\documentclass[%
superscriptaddress,
preprint,
 amsmath,amssymb,
 aps,
prb,
]{revtex4-2}

\usepackage{graphicx}
\usepackage{dcolumn}
\usepackage{bm}

\usepackage{pifont} 
\usepackage{xcolor} 
\usepackage{makecell}
\usepackage[colorlinks=true,citecolor=blue,linkcolor=blue,pdfhighlight =/O]{hyperref}

\newcommand{\eref}[1]{Eq.~(\ref{#1})}

\newcommand{\sfref}[1]{Figure~\ref{#1}}
\newcommand{\sfrefs}[1]{Figures~\ref{#1}}
\newcommand{\fref}[1]{Fig.~\ref{#1}}
\newcommand{\frefs}[1]{Figs.~\ref{#1}}

\hypersetup{urlcolor=blue}

\begin{document}


\title{Fabry-Perot Interferometric Calibration of 2D Nanomechanical Plate Resonators}

\author{Myrron Albert Callera Aguila}
\email{maguila@gate.sinica.edu.tw}
\affiliation{Department of Engineering and System Science, National Tsing Hua University, Hsinchu, 30013, Taiwan}
\affiliation{Nano Science and Technology Program, Taiwan International Graduate Program, National Tsing Hua University and Academia Sinica}
\affiliation{Institute of Physics, Academia Sinica, Nankang, Taipei, 11529, Taiwan}

\author{Joshoua Condicion Esmenda}%
\email{jesmenda@gate.sinica.edu.tw}
\affiliation{Department of Engineering and System Science, National Tsing Hua University, Hsinchu, 30013, Taiwan}
\affiliation{Nano Science and Technology Program, Taiwan International Graduate Program, National Tsing Hua University and Academia Sinica}
\affiliation{Institute of Physics, Academia Sinica, Nankang, Taipei, 11529, Taiwan}

\author{Jyh-Yang Wang}
\affiliation{Institute of Physics, Academia Sinica, Nankang, Taipei, 11529, Taiwan}

\author{Teik-Hui Lee}
\affiliation{Institute of Physics, Academia Sinica, Nankang, Taipei, 11529, Taiwan}

\author{Chi-Yuan Yang}
\affiliation{Institute of Physics, Academia Sinica, Nankang, Taipei, 11529, Taiwan}

\author{Kung-Hsuan Lin}
\affiliation{Institute of Physics, Academia Sinica, Nankang, Taipei, 11529, Taiwan}

\author{Kuei-Shu Chang-Liao}
\affiliation{Department of Engineering and System Science, National Tsing Hua University, Hsinchu, 30013, Taiwan}

\author{Sergey Kafanov}
\affiliation{Department of Physics, Lancaster University, Lancaster, LA1 4YB, United Kingdom}

\author{Yuri A. Pashkin}
\affiliation{Department of Physics, Lancaster University, Lancaster, LA1 4YB, United Kingdom}

\author{Chii-Dong Chen}
\email{chiidong@phys.sinica.edu.tw}
\affiliation{Institute of Physics, Academia Sinica, Nankang, Taipei, 11529, Taiwan}

\date{\today}

\begin{abstract}
Displacement calibration of nanomechanical plate resonators presents a challenging task. Large nanomechanical resonator thickness reduces the amplitude of the resonator motion due to its increased spring constant and mass, and its unique reflectance. Here, we show that the plate thickness, resonator gap height, and motional amplitude of circular and elliptical drum resonators, can be determined in-situ by exploiting the fundamental interference phenomenon in Fabry-Perot cavities. The proposed calibration scheme uses optical contrasts to uncover thickness and spacer height profiles, and reuse the results to convert the photodetector signal to the displacement of drumheads that are electromotively driven in their linear regime. Calibrated frequency response and spatial mode maps enable extraction of the modal radius, effective mass, effective driving force and Young’s elastic modulus of the drumhead material. This scheme is applicable to any configuration of Fabry-Perot cavities, including plate and membrane resonators. 
\end{abstract}

\maketitle


Nanomechanical resonators (NMRs) are exceptional force and mass sensors\cite{Ekinci2005NEMS,Imboden2014}, which made them valuable test platforms for the investigation of various phenomena at the nanoscale such as synchronization\cite{Fon2017Complex,Matheny2019Exotic}, noise\cite{Cleland2002Noise,Maillet2017Freq}, nonlinearity\cite{davidovikj2017nonlinear,Yang2019SpaMod}, and light-matter interaction\cite{thompson2008strong,andrews2014bidirectional,bagci2014optical}. NMRs with flexural modes (i.e. plates and beams) have attracted interest due to their linear response even to large deformation-inducing forces\cite{CGomez2012Elastic,wong2010characterization}. The well-known mechanical properties of the bulk material and its geometry determine plate and beam frequencies that can be predicted to a high accuracy. This allows the realization of unique device applications such as nanomechanical mass spectrometers\cite{Naik2009Mass,Sader2018Mass}, phononic crystals built from NMR arrays\cite{cha2018elastic,WangY2019HBN}, and complex networks of NMRs embedded in electrical circuits\cite{Fon2017Complex,Matheny2019Exotic}, and cavity-mediated quantum systems\cite{Gartner2019SiN}.

Calibration of NMR displacement is important for quantification of device characteristics in sensing applications. While the spatial dynamics of membrane NMRs, whose behavior is to a large extent determined by tensile stress, have been investigated in great detail with optical interferometry\cite{de2016tunable,davidovikj2016visualizing,Kim2018,Yang2019SpaMod}, plate NMRs have been less explored. Studies involving Fabry-Perot (FP) cavities have introduced calibration of the vibrational amplitudes of membrane and string NMRs\cite{HAUER2013,davidovikj2016visualizing,dolleman2017amplitude,de2019absolute}. However,  they  are  hardly  applicable  to  plate  NMRs  because  of  reduced  vibrational amplitudes owing to the increased spring constant and mass, and unique reflectance versus FP cavity length arising from thick absorptive  materials  such  as  niobium  diselenide. Also, there are cases where a smaller spacing is preferred over the optimal spacing for interferometric detection. These cases include mechanical frequency tuning by low gate voltages\cite{sazonova2004tunable,chen2013graphene,Tsoukalas2020Trans}, and stronger electomechanical coupling between mechanical resonators and microwave cavities\cite{weber2016force,luo2018phonon}.

In this Letter, we show that the motion of plate NMRs can be calibrated by considering multilayer wave interferences occurring on the FP structure. To demonstrate the robustness of the technique, a thick 2D material, NbSe$_{2}$, is used as the drumhead. This approach allows determination of the layer thickness, spacer height and device responsivity of each translucent flexible mirror. Our calibration scheme reveals subnanometer mechanical displacements for the measured response of plate NMRs with thickness exceeding 50 nm.

\begin{figure}[t]
    \begin{center}
\includegraphics[width=8.6cm]{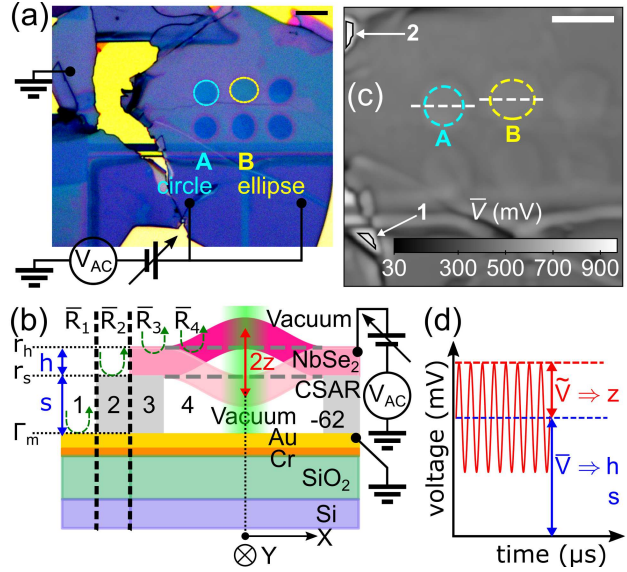}
\caption{(a) Optical micrograph of the NbSe$_{2}$ plate resonator devices. The circuit diagram shows the electromotive scheme used in driving the resonator. (b) Schematic drawing of the optical cross-section of the device (zones 3 and 4) and the references (zones 1 and 2) as measured via Fabry-Perot interference. The actuation circuit is added for clarity. (c) Confocal image showing devices A and B as scanned with a probe laser beam with wavelength $\lambda=532\,$nm. Scale bars in (a) and (c) are set to $10\,\mu$m. (d) Sketch of the output voltage of the fast photodetector versus time of a driven flexural resonator.}
    \label{fig:device_scheme}
    \end{center}
\end{figure}

\sfref{fig:device_scheme}(a) shows Device A, a circular plate with a hole diameter of 7$\,\mu$m, and Device B, an elliptical plate with hole diameters of 8$\,\mu$m (X, major axis) and 7$\,\mu$m (Y, minor axis). The devices share the same flake, ground electrode, and driving voltages. The NbSe$_{2}$ flake and ground electrodes are separated by the insulating layer (electron-beam resist CSAR-62) and vacuum spacers, and hence form FP cavities for detection, and capacitors for actuation. A large rectangular opening, located tens of microns below the drum centers, allows the flake to connect to the voltage-controlled Au/Cr electrodes. The motion of the electromotively driven plates is detected interferometrically in a high vacuum environment\cite{SupMat}. 

Our method relies on different contrast of light elastically reflected from each zone, as shown in \fref{fig:device_scheme}(b). The flake (pink bar) acts as a translucent movable mirror with thickness $h$, which is separated from the ground electrode by a spacer of height $s$. For convenience, the reflected intensity is expressed in terms of the reflectance $R$, which is the ratio of the total reflected light intensity to the incident intensity. Stationary mirrors have only DC component $R=\overline{R}$ while movable mirrors have both $\overline{R}$ and AC component $\widetilde{R}$.  Zones 1 and 2 represent two stationary mirrors: stacks of gold, orange, green and blue bars having reflectance $R_{1}=\overline{R}_{1}$ and a mirror covered with a spacer (light gray) having reflectance $R_{2}=\overline{R}_{2}$, respectively. Zone 3 represents two stationary mirrors separated by a dielectric gap (clamp) with reflectance $R_{3}=\overline{R}_{3}$. Finally, zone 4 is the main FP cavity composed of one stationary and one movable mirrors, which are separated by a vacuum gap with reflectance $R_{4}$. Here, zones 1 and 2 are references for zones 4 and 3, respectively. Scanning mirrors in the measurement setup move the laser spot in each zone a distance $X$ and $Y$ away from the drums' center. 

Application of DC and AC voltages to the flake exerts an attractive force; the NMR responds with an out-of-plane motional amplitude $z$ at a driving frequency $f_{d}$. Due to the position and motion of the movable mirror in zone 4, the main FP cavity has reflectance $R_{4}=\overline{R}_{4}+\widetilde{R}_{4}(f_{d})$, with $\overline{R}_{4}\gg\widetilde{R}_{4}(f_{d})$. \sfref{fig:device_scheme}(d) shows the photodetector output signal $V$ acquired from $R_{4}$. Both the DC component $\overline{V}$ and the AC component $\widetilde{V}$ of the output signal are proportional to $\overline{R}_{4}$ and $\widetilde{R}_{4}$, respectively. Amplitude $z$ is determined after obtaining $h$ and $s$. 

Though we calculate $\overline{R}_{1-4}$ using the multilayer interference approach\cite{Golla2013Optical,orfanidis_2016,chen2018vibration} (MIA), the reflectance of FP cavities with four interfaces\cite{jung2007simple,casiraghi2007rayleigh} $\overline{R}_{3,4}$ captures the stationary reflections occurring for each drum. Here, we assume that the coherent probe light, having wavelength $\lambda$, originates from a point source and propagates from a semi-infinite vacuum layer. The drum and the bottom mirror have complex refractive indices $\hat{n}_{h}$\cite{hill2018comprehensive} and $\hat{n}_{m}$, respectively, whereas the spacers have real refractive index $\hat{n}_{s}$ ($\hat{n}_{s,drum}$ for the vacuum spacer and $\hat{n}_{s,clamp}$ for the CSAR-62 spacer). In this geometry, the vacuum-NMR, NMR-spacer, and spacer-mirror interfaces contribute significantly to the cavity’s overall reflectance. The total reflectance is defined as
\begin{equation}
    {\overline{R}_{3,4}} = {\left| {\frac{{{r_{h}} + {r_{s}}{e^{ - 2j{\delta _{h}}}} + \left[ {{r_{h}}{r_{s}} + {e^{ - 2j{\delta _{h}}}}} \right]{\Gamma _{m}}{e^{ - 2j{\delta _{s}}}}}}{{1 + {r_{h}}{r_{s}}{e^{ - 2j{\delta _{h}}}} + \left[ {{r_{s}} + {r_{h}}{e^{ - 2j{\delta _{h}}}}} \right]{\Gamma _{m}}{e^{ - 2j{\delta _{s}}}}}}} \right|^2}
    \label{eqn:Rbar}
\end{equation}
where $\delta_{h}=2\pi \hat n_{h}h/\lambda$ is the optical phase thickness of the NMR, $\delta_{s}=2\pi \hat n_{s}s/\lambda$ is the optical phase thickness of the spacer, $r_{h}$=$(1- \hat n_{h})/(1+ \hat n_{h})$ is the Fresnel coefficient of the vacuum-NMR interface, $r_{s}$=$(\hat n_{h}- \hat n_{s})/(\hat n_{h} + \hat n_{s})$ is the Fresnel coefficient of the NMR-spacer interface, and $\Gamma_{m}=(\hat n_{s}- \hat n_{m})/(\hat n_{s}+ \hat n_{m})$ is the equivalent Fresnel coefficient of the spacer-mirror interface. For convenience, $\Gamma_{m}$ is computed using MIA\cite{SupMat}. 

\sfref{fig:device_scheme}(c) shows the topographical features of the drum devices as probed by a continuous wave laser beam. Apparently, the reflectance signal measured along the white dashed lines drawn across devices A and B contains $\overline{V}_{3}$, taken outside the dashed ellipses, and $\overline{V}_{4}$, taken within the dashed ellipses. Polygons 1 and 2 give average values of $\overline{V}_{1}$ and $\overline{V}_{2}$. The colored dashed ellipses, representing the hole diameters measured in \fref{fig:device_scheme}(a), are smaller than the light gray ellipses. These gray ellipses manifest in \fref{fig:device_scheme}(a) as concentric purple rings seen for each drum. These features arise when the flake transfer, that is based on elastomeric stamps, deforms the edge of every drum hole. 
\begin{figure}[htbp]
    \begin{center}
\includegraphics[width=8.6cm]{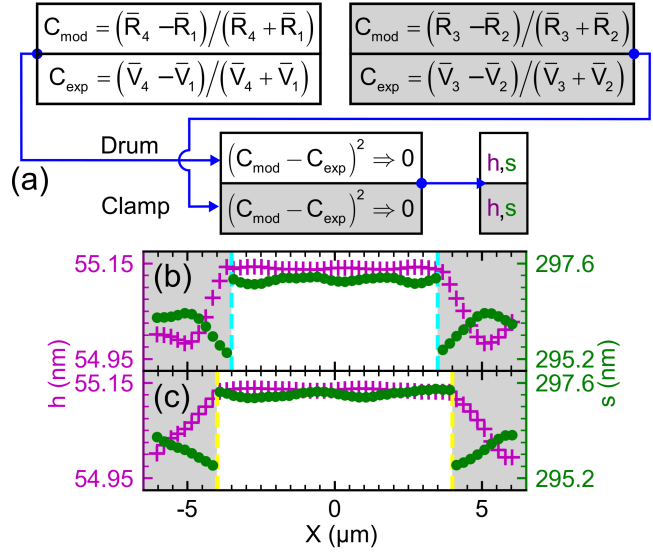}
\caption{(a) Diagram for determining $h$ and $s$ for the clamp and drum zones. Minimization of the difference between the experimental contrast ($C_{\mathrm{exp}}$) and the modelled contrast ($C_{\mathrm{mod}}$) results in $h$ and $s$ profiles for Device A (b) and Device B (c). Colored dashed lines refer to the hole radius set in \frefs{fig:device_scheme}(a,c), separating the drum (white fills) and clamp (gray fills) zones.}
    \label{fig:prof_scheme}
    \end{center}
\end{figure}

Since $\overline{R}$ is susceptible to scattering losses\cite{Liu2013InAs}, we circumvent this issue by normalizing the Michelson contrast\cite{jung2007simple} of each FP cavity to their reference. Having defined the experimental and calculated reflectance, the cavity’s optical contrast, $C$, is quantified as $C=(\overline{R}_{\mathrm{3,4}}-\overline{R}_{\mathrm{2,1}})/(\overline{R}_{\mathrm{3,4}}+\overline{R}_{\mathrm{2,1}})$, where $\overline{R}_{\mathrm{3,4}}$ is the stationary reflectance of the FP cavity, and $\overline{R}_{\mathrm{2,1}}$ is the stationary reflectance of the cavity’s reference. Apparently, $C$ ranges between -1 and 1, with zero denoting no difference with the reference. If $C$ is positive, then the cavity is brighter than the reference. Otherwise, the cavity is darker than its reference. 

The output voltages measured for each pixel along the dashed lines in \fref{fig:device_scheme}(c) are converted into contrast values for devices A and B, as depicted in \fref{fig:prof_scheme}(a). The experimental contrast $C_{\mathrm{exp}}$ represents the ratio of voltages acquired from different zones in the confocal image of each device while the modelled contrast $C_{\mathrm{mod}}$ is derived using MIA\cite{SupMat}. \sfrefs{fig:prof_scheme}(b-c) show the resulting $h$ and $s$ cross-section profiles acquired from minimizing the difference between the experimental contrast values and the contrasts generated by MIA. The mean plate thicknesses and spacer heights for the two devices are in excellent agreement with the mean values listed in Table \ref{tab:table1}. The spacer height for both drums and clamps agrees well with the stylus profilometer measurements. From the flake thickness of about 55$\,$nm, we deduce 92 layers of NbSe$_{2}$ sheets assuming a single layer thickness of $0.6\,$nm\cite{castellanos2010optical}.

\begin{table}[htb]
\caption{%
Mean flake and spacer thicknesses of NbSe$_{2}$ drum and clamp zones
}
\begin{ruledtabular}
\begin{tabular}{ccc}
\textrm{Devices}&
\textrm{A}&
\textrm{B}\\
\colrule
$h_{drum}$ (nm) & 55.139 $\pm$ 0.002 & 55.135 $\pm$ 0.002\\
$s_{clamp}$ (nm) & 55.03 $\pm$ 0.05 & 55.05 $\pm$ 0.04\\
$s_{drum}$ (nm) & 297.2 $\pm$ 0.1 & 297.3 $\pm$ 0.1\\
$s_{drum}$ (nm) & 296.0 $\pm$ 0.3 & 295.9 $\pm$ 0.3\\
\end{tabular}
\end{ruledtabular}
\label{tab:table1}
\end{table}

The $h$ profiles in \frefs{fig:prof_scheme}(b-c) show a hundred picometer variation between the drum and clamp zones. Meanwhile, buckling is observed in the $s$ profiles in \frefs{fig:prof_scheme}(b-c) as $s_{drum}$ for both devices are greater than $s_{clamp}$ by $1.2 - 1.4\,$nm. We see the drumheads bulge\cite{Minot2003Tuning,Zheng2017Hexagonal} presumably due to the pressure difference between the trapped air in the drum hole and the vacuum environment. The surface roughness of the movable mirror likely comes from the thermally-grown oxide\cite{Blasco2001afm} on the surface of the stationary mirror.

Having determined the mean $h_{drum}$ and $s_{drum}$, we evaluate the optical reflectance-to-displacement responsivity $\left| d\overline{R}_{4}/ds \right|$ of each drum. This quantity is obtained from $\widetilde{R}_{4} (f_{d})$ = $\left| d\overline{R}_{4}/ds \right| z(f_{d})$. The AC component reflected from zone 4 and characterized by $\widetilde{R}_{4}$, being purely due to mechanical motion, is insensitive to any scattering losses as this wave is amplitude-modulated. \eref{eqn:Rbar} is then corrected by a prefactor that accounts for the finite spot size of the probe Gaussian beam\cite{SupMat}.  

\begin{figure}[htbp]
\includegraphics[width=8.6cm]{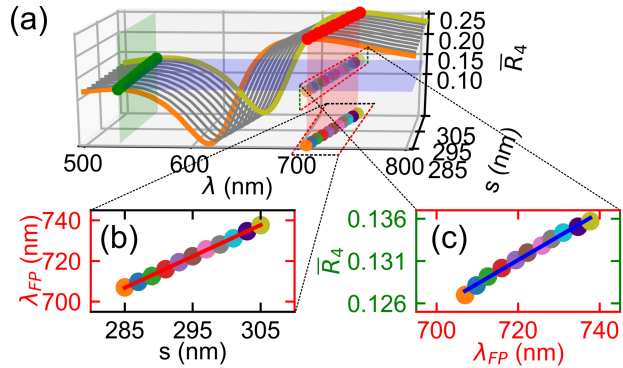}
\begin{center}
\caption{(a) Waterfall plot of FP reflectivity as a function of $\lambda$ at varying $s$, with the used probe wavelength (green plane) situated at $\lambda=532\,$nm. (b) Colored scatter plot of $\lambda_{FP}$ as a function of $s$. The slope of the red solid line originates from the intersection of the red plane with the $\lambda-s$ plane in (a). (c) Colored scatter plot of $\overline{R}_{4}$ as a function of the cavity shift $\lambda_{FP}$. The blue solid line comes from the the intersection of the blue plane with the red plane in (a).}
\label{fig:dev_response}
\end{center}
\end{figure}

We define the average $\left| d\overline{R}_{4}/ds \right|$ to account for spatial variations in $s_{drum}$ across the plate due to the pressure difference and DC voltage. Note that each complex-valued refractive index is dependent on the probing wavelength; this translates into the wavelength-dependent $r_{h}$, $r_{s}$ and $\Gamma_{m}$. We modeled $\left| d\overline{R}_{4}/s \right|_{avg}$ by the chain rule $\left| {\Delta {{\bar R}_4}\left( \lambda  \right)}/{\Delta {\lambda_{FP}}} \right| {\left| {\Delta {\lambda_{FP}}} / {\Delta s} \right|_{s=s_{drum}}}$, where $\Delta\overline{R}_{4}/\Delta\lambda_{FP}$ is the change of $\overline{R}_{4}$ with regards to the wavelength shift in the FP cavity, and $\Delta\lambda_{FP}/\Delta s$ is the wavelength shift of the FP cavity caused by the change of the spacer gap. 
The resulting dependences are shown as a waterfall plot in \fref{fig:dev_response}(a) with a gap range exceeding the uncertainty of our stylus profilometer\cite{SupMat}. \sfref{fig:dev_response}(a) demonstrates larger $\overline{R}_{4}$ at near-infrared wavelengths. \sfref{fig:dev_response}(b) shows the peak wavelength of the cavity, falling in the near-infrared range, shifting linearly with a slope of $1.543\,$nm/nm as $s_{drum}$ increases from $285\,$nm to $305\,$nm. \sfref{fig:dev_response}(c) shows how the shift consequently increases $\overline{R}_{4} \left(\lambda \right)$ linearly, with a slope of $0.28\times10^{-6}\,$nm$^{-1}$. The product of these two slopes, $ \left| d\overline{R}_{4}/ds \right|_{avg}=0.43 \times 10^{-3}\,$nm$^{-1}$, agrees with $\left| d\overline{R}_{4}/ds \right|_{s=s_{drum}}=0.40\times 10^{-3}\,$nm$^{-1}$ that is evaluated from the gradient of the $\overline{R}_{4}$ with respect to $s$\cite{SupMat}. The linear behavior seen in \frefs{fig:dev_response}(b-c) is in contrast to the non-linear dependence observed for optically-thin membranes in the same ranges of $s$\cite{SupMat}. 

\begin{figure}[htbp]
\includegraphics[width=8.6cm]{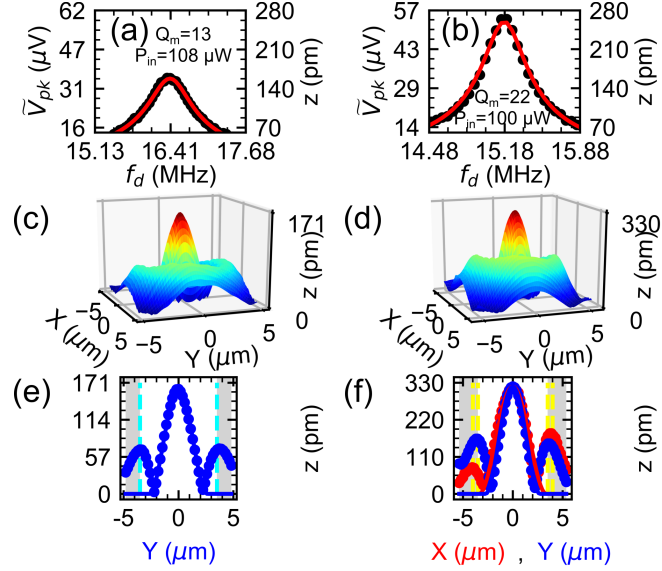}
\caption{Spot-based displacement amplitude response (\color{black}\ding{108}\color{black}) of device A (a) and device B (b) at $V_{DC}$ = 4 V and $V_{AC}$ = 125 mV and their driven resonator fits (red solid lines). The data shown in (c) and (d) refer to the spatial amplitude mode shape of devices A and B at $f_{d}=f_{m}$. The amplitude profile along X (\color{red}\ding{108}\color{black}) and Y (\color{blue}\ding{108}\color{black}) axes of the mode shapes for devices A (e) and B (f) is fitted with a clamped circular plate model (red and blue solid lines for X and Y, respectively). Dashed vertical lines indicate the edges of the holes.}
\label{fig:cal_NMRs} 
\end{figure}

We use the average responsivity together with the interferometer system gain $S$($\lambda$) (V/W), and the laser probe power $P_{in}$ to define the displacement amplitude $z$ as   
\begin{equation}
z\left( {{f_d},X,Y} \right) = \frac{{\widetilde V_{pk}\left( {{f_d},X,Y} \right)}}{{\left| d\overline{R}_{4}\left( {{\lambda},{h_{drum}},{s_{drum}}} \right)/ds \right| S\left( {{\lambda}} \right){P_{in}}}}
\label{eqn:z_convert}
\end{equation}
where $\widetilde{V}_{pk}$ is the frequency and position-dependent peak voltage response of the NMR. The denominator in \eref{eqn:z_convert}, when squared, represents the transduction factor $\alpha$ (in V$^2$/m$^2$) that one can deduce from the measured Brownian motion of a mechanical resonator probed by an interferometric system\cite{HAUER2013,davidovikj2016visualizing}. This quantity accounts for the device responsivity and the detection parameters in the interferometric setup\cite{SupMat}. We deduce transduction factors of $0.20\,\mu$V/pm for device A, and $0.22\,\mu$V/pm for device B for probe powers listed in \frefs{fig:cal_NMRs}(a-b).

\sfrefs{fig:cal_NMRs}(a-b) show the measured voltage response and its corresponding $z$ for devices A and B. The measured $z$ response profile agrees well with a driven mechanical resonator model in the linear regime\cite{Schmid2016Fund}:
\begin{equation}
z\left( {{f_d},X,Y} \right) = \frac{{{A_{eff}}}}{{4{\pi ^2}\sqrt {{{\left( {f_d^2 - f_m^2} \right)}^2} + {{\left( {{{{f_d}{f_m}} \mathord{\left/
 {\vphantom {{{f_d}{f_m}} {{Q_m}}}} \right.
 \kern-\nulldelimiterspace} {{Q_m}}}} \right)}^2}} }} \Phi \left( {X,Y} \right)
 \label{eq:linear_res}
\end{equation}
where $f_{m}$ is the fundamental mode frequency, $Q_{m}$ is the mode quality factor, and $A_{eff}$ is the amplitude expressed as effective acceleration. $\Phi(X,Y)$ is the mode shape of the plate described as
\begin{equation}
\Phi \left( {X,Y} \right) = {K_0}\left[ {{J_0}\left( {\beta k(X,Y)} \right) - \frac{{{J_0}\left( \beta  \right)}}{{{I_0}\left( \beta  \right)}}{I_0}\left( {\beta k\left( {X,Y} \right)} \right)} \right]
\label{eq:def_surf}
\end{equation}
where $J_{0}$ and $I_{0}$ are the zeroth Bessel and modified Bessel functions of the first kind, respectively, $\beta = 3.1961$ is the fundamental mode constant for a clamped circular plate, and $K_{0} = 0.947$ is a normalization constant. Here, $k\left( {X,Y} \right) = \sqrt {{{\left( {{X \mathord{\left/
 {\vphantom {X a}} \right.
 \kern-\nulldelimiterspace} a}} \right)}^2} + {{\left( {Y/b} \right)}^2}}
$ represents the normalized coordinates away from the maximum of $z$, where $a$ and $b$ represent the NMR major (X axis) and minor (Y axis) modal radii, respectively. By setting $\Phi(X,Y)=1$, we measure $z_{A}=158\,\pm\,2\,$pm for device A, and $z_{B}=259\,\pm\,3\,$pm for device B. Their magnitudes are three orders of magnitude smaller than $h_{drum}$ and $s_{drum}$. 

\begin{table}[htb]
\caption{%
Modal properties of NbSe$_{2}$ devices
}
\begin{ruledtabular}
\begin{tabular}{cccc}
\textrm{Specifications}&
\textrm{Device A}&
\textrm{Device B}&
\textrm{Method}\\
\colrule
$z_{max}$ (pm) & 161 & 320 & \eref{eq:def_surf} \\
$a$ ($\mu$m) & 2.7 $\pm$ 0.2 & 3.19 $\pm$ 0.06 & \eref{eq:def_surf}\\
$b$ ($\mu$m) & 2.6 $\pm$ 0.2 & 2.66 $\pm$ 0.02 & \eref{eq:def_surf} \\
$m_{eff}$ (fg) & 1.4 & 1.74 & \cite{SupMat} \\
$A_{eff}$ (km/s$^{2}$) & 132 & 132 & \eref{eq:linear_res} \\
$F_{eff}$ (pN) & 191 & 229 & \cite{SupMat} \\
$E_{Y}$ (GPa) & \multicolumn{2}{c}{135 $\pm$ 13} & \cite{SupMat}\\
\end{tabular}
\end{ruledtabular}
\label{tab:table2}
\end{table}

By driving the plates at $f_{m}$, and probing their spatial mode shape with scanning mirrors, we observe surface plots of $z$ for devices A and B as shown in \frefs{fig:cal_NMRs}(c-d). \sfrefs{fig:cal_NMRs}(e-f) show X and Y axes cuts, with both axes intersecting at $z_{max}$ of \frefs{fig:cal_NMRs}(c-d). They reveal $z$ profiles that agree with \eref{eq:def_surf}, with $a$ and $b$ acting as free parameters. $z_{max}$, $a$, and $b$ of the two plates are listed in Table \ref{tab:table2}. The discrepancy in the values of $z_{B}$ and $z_{max}$ of device B is due to the location of the laser spot that probed \fref{fig:cal_NMRs}(b). Whereas $z_{A}$ lies at $X=Y\approx\,$0, $z_{B}$ lies at $X,Y\approx1\,\mu$m away from the spatial peak. Both $a$ and $b$ for devices A and B are smaller than the hole radii (set as cyan and yellow dashed lines in \frefs{fig:cal_NMRs}(e-f)), making $f_{m}$ for both devices higher than the designed values. Moreover, device A shows unexpected elliptical modal behavior with a miniscule difference between $a$ and $b$, which is due to the fabrication process. 

Table \ref{tab:table2} lists other NMR-related quantities that are derived from Figure 4 such as the effective mass $m_{eff}$, acceleration $A_{eff}$, force $F_{eff}$, and Young’s elastic modulus $E_{Y}$. These quantities are derived from a clamped elliptical plate model\cite{SupMat}. The estimated $E_{Y}$ is within the range of reported values for bulk NbSe$_{2}$ flakes\cite{Barmatz1975,sengupta2010electromechanical}. These quantities are obtained without inducing damage on the flake, and are independent of the actuation scheme. 

\eref{eq:def_surf} does not explain the asymmetric sinusoidal waves propagating beyond the drum edges seen in \frefs{fig:cal_NMRs}(e-f). These waves are signatures of support losses due to imperfect flake clamping at the edges\cite{Pandey2009anchor}. Discussing the waves’ origin goes beyond the scope of this study, though resolving the waves' amplitude, which is $1/3$ of $z_{max}$, demonstrate the capability of our method to visualize acoustic waves in NMRs \cite{WangY2019HBN}.

In summary, we demonstrated an in-situ, non-invasive method of calibrating motional displacement of plate NMRs by exploiting wave interference phenomena in FP cavities.  Using a probe laser beam, and applying MIA to different realizations, we determine cross-sectional profiles of the NMR thickness and spacer height, transduction factors of NbSe$_{2}$ plate resonators, and subnanometer motional amplitudes that helped examine modal properties of the drumheads. We foresee that this method will be applicable to flexural and acoustic wave resonators of various geometries. 

\begin{acknowledgments}
We acknowledge the contributions of T.-H. Hsu and W.-H. Chang in fabricating devices and in building the experimental setup. We thank A.F. Rigosi for sharing the measured dielectric constant spectra of bulk and few-layers of NbSe$_{2}$. We thank the Taiwan International Graduate Program for the financial support. This project is funded by Academia Sinica Grand Challenge Seed Program (AS-GC-109-08), Ministry of Science and Technology (MOST) of Taiwan (107-2112-M-001-001-MY3), Cost Share Programme (107-2911-I-001-511), the Royal Society International Exchanges Scheme (grant IES$\backslash$R3$\backslash$170029), and iMATE(2391-107-3001). We extend our gratitude for the Academia Sinica Nanocore Facility.

\textit{Attributions.-}M.A.C.A. and J.C.E. contributed equally in this work. C.-D.C. conceived the devices and supervised the project; J.C.E. fabricated the devices. M.A.C.A. and J.-Y.W. modeled the calibration scheme. C.-Y.Y. and K.-H.L. designed and built the setup for optical measurements. M.A.C.A., J.C.E. and C.-Y.Y. performed the experiment. M.A.C.A., J.C.E., J.-Y.W., T.-H.L., K.-S.C.-L., S.K., Y.P. and C.-D.C. analyzed the data, performed simulations and wrote the manuscript; all authors discussed the results and contributed to the manuscript.
\end{acknowledgments}



\providecommand{\noopsort}[1]{}\providecommand{\singleletter}[1]{#1}%

\end{document}